\documentclass[12pt]{article}
\usepackage{epsfig}

\newcommand{\nn}{\nonumber}

\def\gev{~{\rm GeV}}
\def\ale{\alpha_{\rm elm}}
\def\als{\alpha_{\rm s}}
\def\bbb{B\overline{B}}
\def\qbq{q\overline{q}}

\newcommand{\tr}[1]{{\bf #1}_\perp}
\newcommand{\ov}[1]{\overline#1} 
\newcommand{\half}{{\textstyle \frac{1}{2}}}

\begin{document}
\thispagestyle{empty}
\begin{flushright}
WU B 02-06 \\
NORDITA-2002-15 HE \\
hep-ph/0206288\\
October 2002\\[5em]
\end{flushright}

\begin{center}
\end{center}
\begin{center}

{\Large\bf Two-Photon Annihilation into Baryon-Antibaryon Pairs} \\
\vskip 3\baselineskip

M.\ Diehl\,\footnote{Email: mdiehl@physik.rwth-aachen.de}
\\[0.5em]
{\small {\it Institut f\"ur Theoretische Physik E, RWTH Aachen, 52056
Aachen, Germany}}\\
\vskip\baselineskip

P.\ Kroll\,\footnote{Email: kroll@physik.uni-wuppertal.de}
\\[0.5em]
{\small {\it Fachbereich Physik, Universit\"at Wuppertal, 42097
Wuppertal, Germany}}\\
\vskip \baselineskip

and C.\ Vogt\,\footnote{Email: cvogt@nordita.dk}
\\[0.5em]
{\small {\it Nordita, Blegdamsvej 17, 2100 Copenhagen, Denmark}}\\
\vskip \baselineskip

\end{center}

\vskip 3\baselineskip

\begin{abstract}
We study the handbag contribution to two-photon annihilation into
baryon-antiba-ry\-on pairs at large energy and momentum transfer.  We
derive factorization of the process amplitude into a hard
$\gamma\gamma \to \qbq$ subprocess and form factors describing the
soft $\qbq\to\bbb$ transition, assuming that the process is dominated
by configurations where the (anti)quark approximately carries the full
momentum of the (anti)baryon. The form factors represent moments of
time-like generalized parton distributions, so-called $\bbb$
distribution amplitudes.  A characteristic feature of the handbag
mechanism is the absence of isospin-two components in the final state,
which in combination with flavor symmetry provides relations among the
form factors for the members of the lowest-lying baryon octet.
Assuming dominance of the handbag contribution, we can describe
current experimental data with form factors of plausible size, and
predict the cross sections of presently unmeasured $\bbb$ channels.
\end{abstract}

\newpage

\section{Introduction}
\begin{figure}[ht]
\begin{center}
\psfig{file=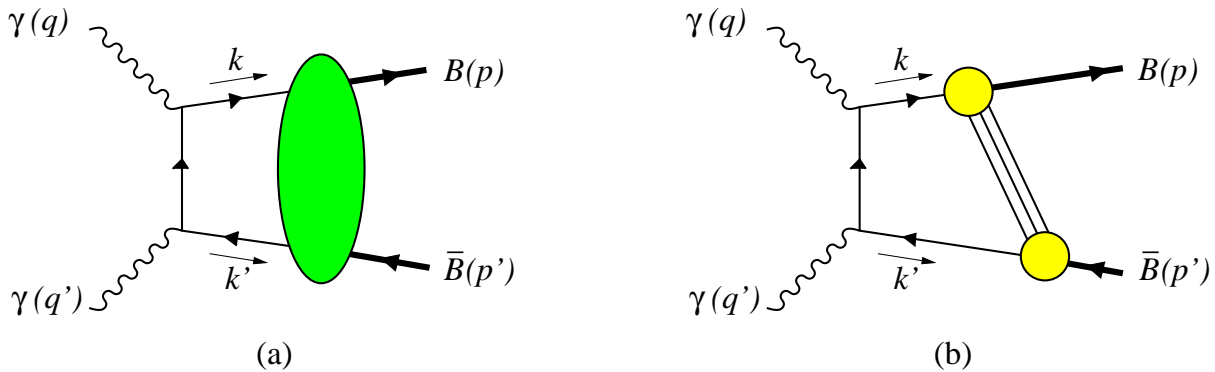,width=0.95\textwidth}
\caption{{}(a) Handbag factorization of baryon pair production
$\gamma\gamma\to \bbb$ at large $s$, $t$, and $u$. The hard scattering
subprocess is shown at leading order in $\alpha_s$, and the blob
represents the $\bbb$ distribution amplitude. A second graph is
obtained by interchanging the photon vertices.  (b)~The physical
mechanism of the handbag diagrams.  The quark hadronizes into $B$ and
the antiquark into $\ov{B}$, with any number of soft partons
connecting the two parton-hadron vertices.}
\label{fig:handbag}
\end{center}
\end{figure}

In this article we study the annihilation of two photons into
baryon-antibaryon ($\bbb$) pairs at large Mandelstam variables $s\sim
-t \sim -u$ in the handbag approach recently developed for two-photon
annihilation into pairs of mesons \cite{DKV02}.  As in the meson case,
the handbag amplitude (see Fig.\ \ref{fig:handbag}) factorizes into a
hard $\gamma\gamma\to \qbq$ subprocess and form factors representing
moments of generalized distribution amplitudes
\cite{mue1994,die1998a}. These distribution amplitudes are time-like
versions of generalized parton distributions, which encode the soft
physics information in processes such as deeply
virtual~\cite{ji1996,rad1996} or wide-angle~\cite{rad1998,DFJK1}
Compton scattering.  The latter is in fact related to two-photon
annihilation by crossing.  It is important to realize that, since we
take the $\bbb$ system to have large invariant mass, the $q\ov{q}\to
\bbb$ transition can only be soft if the additional $q\ov{q}$ and
possibly gluon pairs created in the hadronization process have soft
momenta.  In other words, the initial quark and antiquark must each
take approximately the full momentum of one final state
hadron. Compared to the case of mesons, our derivation for baryons
will make the additional, plausible, assumption that the process is
dominated by configurations where it is the quark that approximately
moves in the direction of the baryon, whereas the antiquark
approximately moves in the direction of the antibaryon.  This
assumption is equivalent to the valence quark approximation, widely
used in other contexts.

For both processes, wide-angle Compton scattering off baryons and
two-photon annihilation into $\bbb$ pairs, the handbag contribution
can dominate for large but not asymptotically large Mandelstam
variables.  Asymptotically the leading-twist contribution will take
over, where in contrast to the handbag mechanism all valence quarks of
the involved hadrons participate in the hard scattering~\cite{bl1980}.
The handbag contribution formally represents a power correction to the
leading-twist one. The onset of the leading-twist regime is however
expected to occur for $s$ much larger than experimentally available.
A more detailed discussion of the relation between the leading-twist 
and soft handbag mechanisms, and of other power suppressed 
contributions is given in \cite{DKV02}.

Two-photon annihilation into hadrons pairs has also been studied for
the case where one of the photons has a virtuality $Q^2$ much larger
than the squared invariant mass $s$ of the hadron pair.  In this
kinematics, which is complementary to the one studied in the present
article, handbag factorization of the process amplitude has been shown
to hold for asymptotically large photon virtualities
\cite{mue1994,die1998a,fre1999}.  In other words, the handbag provides
the leading-twist contribution in the limit of large $Q^2$ at fixed
$s$.

Our paper is organized as follows: In Sect.~\ref{sec:gda} we define
the $\bbb$ distribution amplitudes and discuss some of their
properties. Sect.~\ref{sec:handbag} is devoted to the calculation of
the handbag amplitude for $\gamma\gamma\to \bbb$. Flavor symmetry
properties of the handbag amplitude are investigated in
Sect.~\ref{sec:flavor}, and a comparison to experiment is presented in
Sect.~\ref{sec:pheno}. We end our paper with a few concluding remarks.

\section{Distribution amplitudes for baryon-antibaryon pairs}
\label{sec:gda}

Generalized distribution amplitudes for meson pairs have been
discussed in detail in the literature \cite{die1998a,pol1998}.  Here
we introduce their counterparts for baryon-antibaryon pairs and
present some of their general properties.  We use light-cone
coordinates $v = [v^+, v^-, {\bf v}_\perp]$ with $v^\pm = (v^0 \pm
v^3) /\sqrt{2}$ for any four-vector $v$, and define $\bbb$
distribution amplitudes in light-cone gauge by
\begin{eqnarray} 
\lefteqn{P^+ \int \frac{dx^-}{2\pi}\, e^{-i z P^+ x^-}\, 
{}_{\rm out}\big\langle\, B(p\nu) \ov{B}(p'\nu')\,\big|\, 
        \ov{q}(\bar{x})
            \gamma^+ q(0) \,\big|\, 0 \,\big\rangle = } 
\nn\\
&&\qquad\Phi^{q}_{V}(z,\zeta,s)\, \ov{u}(p\nu) \gamma^+ \,v(p'\nu')
       + \Phi^{q}_{S}(z,\zeta,s)\, \frac{P^+}{2m}\, \ov{u}(p\nu) 
                          \,v(p'\nu')\,, 
\nn\\
\lefteqn{P^+ \int \frac{dx^-}{2\pi}\, e^{-i z P^+ x^-}\, 
{}_{\rm out}\big\langle\, B(p\nu) \ov{B}(p'\nu')\,\big|\, 
        \ov{q}(\bar{x}) 
            \gamma^+\gamma_5\, q(0) \,\big|\, 0\, \big\rangle =}  
\nn\\
&&\qquad\Phi^{q}_{A}(z,\zeta,s)\, \ov{u}(p\nu) \gamma^+\gamma_5\, 
                                                            v(p'\nu')
       + \Phi^{q}_{P}(z,\zeta,s)\, \frac{P^+}{2m}\,\ov{u}(p\nu) 
                          \gamma_5\, v(p'\nu')
\label{bbdaV}
\end{eqnarray}
with $\bar{x}=[0, x^-, {\bf 0}_\perp]$.  Here $m$ denotes the mass of
the baryons and $\nu$, $\nu'$ their helicities.  We have further
introduced the sum $P= p+p'$ of the baryon momenta, the invariant mass
$s=P^2$ of the baryon pair, and the skewness
\begin{equation}
   \zeta= p^+ / P^+ .
\label{def-zeta}
\end{equation}
In the following we will also use the notation $\ov{\zeta} = 1-\zeta$.
We have not displayed the dependence of the $\bbb$ distribution
amplitudes on the factorization scale $\mu^2$, which is governed by
the well-known evolution equations for the distribution amplitudes of
a single meson with appropriate quantum numbers.

The $\bbb$ distribution amplitudes are the time-like versions of
generalized parton distributions for a baryon $B$. Let us briefly
comment on the relation of our definitions~(\ref{bbdaV}) with those of
the generalized parton distributions $H$, $E$, $\tilde{H}$,
$\widetilde{E}$, introduced in \cite{ji1996}.  Comparing the Lorentz
structures that multiply the distributions and taking into account
that $p'^+$ turns into $-p'^+$ under crossing of $\ov{B}(p'\nu')$ to
$B(p'\nu')$, we recognize $\Phi_A$ and $\Phi_P$ as the respective
counterparts of $\tilde{H}$ and $\widetilde{E}$.  In the vector
channel one may use the Gordon decomposition
\begin{equation}
\ov{u}(p\nu) \frac{i}{2m}\, 
  \sigma^{+\rho}(p+\varepsilon\, p')_\rho\, v(p'\nu') =  
     \frac12 (1+\varepsilon)\,\ov{u}(p\nu) \gamma^+\, v(p'\nu')
   - \frac{1}{2m}\, (p-\varepsilon\, p')^+\, \ov{u}(p\nu) v(p'\nu')
\label{gordon}
\end{equation}
with $\varepsilon=\pm 1$ to trade the scalar current for the tensor
one.  By crossing the defining relation for $H$ and $E$ one would
obtain the scalar current $\ov{u}(p\nu) \,v(p'\nu')$ multiplied with
$(p'-p)^+ = (1-2\zeta) P^+$ instead of $P^+$.  Defining a distribution
amplitude $\widetilde\Phi_S$ with such a prefactor would however
introduce an artificial singularity of $\widetilde\Phi_S(z,\zeta,s)$
at $\zeta=\frac{1}{2}$, since there is no symmetry by which $(p'-p)^+
\, \widetilde\Phi_S(z,\zeta,s) = P^+ \, \Phi_S(z,\zeta,s)$ has to
vanish at $p^+ = p'^+$.  This is in contrast to the case of the
generalized parton distribution $\widetilde{E}$, where due to time
reversal invariance the product $(p'-p)^+ \widetilde{E}$ occurring in
its definition is zero for $p^+ = p'^+$.

Integrating (\ref{bbdaV}) over $z$ reduces the bilocal $\bbb$ matrix
elements to local ones.  In analogy to the space-like case we obtain a
set of sum rules,
\begin{eqnarray}
F^{q}_{i}(s) &=& \int_0^1 dz\; \Phi^{q}_{i}\,(z,\zeta,s)
\qquad \mbox{for~} i=V,A,P\,, 
\nn\\
(1-2\zeta)\, F^{q}_{S}(s) &=& \int_0^1 dz\; 
                                      \Phi^{q}_{S}\,(z,\zeta,s)
\label{sumrules}  
\end{eqnarray}
with time-like form factors defined as
\begin{eqnarray} 
\label{timelike-ff}
{}_{\rm out}\big\langle\, B(p\nu) \ov{B}(p'\nu')\,\big|\, \ov{q}(0)
                   \gamma^\rho q(0) \,\big|\, 0 \,\big\rangle &=& 
         F^{q}_{V}\, \ov{u}(p\nu) \gamma^\rho \,v(p'\nu')
       + F^{q}_{S}\, \frac{(p'-p)^\rho}{2m}\, \ov{u}(p\nu) 
                          \,v(p'\nu')\,, 
\\
{}_{\rm out}\big\langle\, B(p\nu) \ov{B}(p'\nu')\,\big|\, \ov{q}(0) 
\gamma^\rho\gamma_5\, q(0) \,\big|\, 0\, \big\rangle &=&  
         F^{q}_{A}\, \ov{u}(p\nu) \gamma^\rho\gamma_5\, v(p'\nu')
       + F^{q}_{P}\, \frac{(p'+p)^\rho}{2m}\,\ov{u}(p\nu) 
                          \gamma_5\, v(p'\nu')
\nn
\end{eqnarray}
for each flavor.  Appropriate combinations give the form factors of
the electromagnetic and weak currents, for instance the magnetic and
Pauli form factors
\begin{equation}
G_M(s) = \sum_q \,e_q \, F^{q}_{V}(s) , \qquad
F_2(s) = \sum_q \,e_q \, F^{q}_{S}(s) .
\label{gm}
\end{equation}
The relations (\ref{sumrules}) are valid for any physical value of the
skewness $\zeta$.  They also hold for any value of the factorization
scale $\mu^2$ of the distribution amplitudes, since the vector and
axial vector currents have zero anomalous dimension and the form
factors $F_{V,S}$ and $F_{A,P}$ are scale independent.  Taking higher
moments in $(2z-1)$ leads to $\mu^2$ dependent form factors of local
operators with derivatives, multiplied by polynomials in $(2\zeta-1)$.

{}From charge-conjugation invariance we find the symmetry relations
\begin{eqnarray}
\Phi^{q}_{i}(z,\zeta,s) &=& \hspace{3ex}
                                \Phi^{q}_{i}(\ov{z},\ov{\zeta},s)
\qquad \mbox{for~} i=V,A,P\,, 
\nn\\
\Phi^{q}_{S}(z,\zeta,s) &=& {}- \Phi^{q}_{S}(\ov{z},\ov{\zeta},s)
\label{symm}
\end{eqnarray}
with $\ov{z}=1-z$.  For $\zeta=\frac12$ the distribution amplitudes
$\Phi_{V,A,P}$ are hence symmetric under the replacement
$z\leftrightarrow \ov{z}$, while $\Phi_S$ is antisymmetric (but not
zero).  One may consider $\bbb$-states of definite charge-conjugation
symmetry
\begin{equation}
{\big|}\,{C\pm}{\big\rangle} = 
       \frac12 \, {\Big|} B(p,\nu)\, \ov{B}(p'\nu')\,
               \,\mp\, B(p',\nu')\, \ov{B}(p\nu) \Big\rangle\, ,
\label{c-states}
\end{equation}
satisfying
\begin{equation} 
        {\cal C}\, \big|\,{C\pm}\,\big\rangle = 
             \pm \,\big|\,{C\pm}\,\big\rangle \,,
\end{equation}
where the operator ${\cal C}$ implements charge conjugation in Hilbert
space.  Replacing the state $\langle\, B(p,\nu)\, \ov{B}(p'\nu')\,|$
with $\langle\, {C\pm} |$ in the definition (\ref{bbdaV}), we obtain
on its right-hand side the linear combinations
\begin{eqnarray}
\Phi^{q(\pm)}_{i}(z,\zeta,s) &=& \frac12 \Big[ \Phi^{q}_{i}(z,\zeta,s)
                  \mp \Phi^{q}_{i}(\ov{z},{\zeta},s)\Big] 
\qquad \mbox{~for~} i=V,S \,,
\\
\Phi^{q(\pm)}_{i}(z,\zeta,s) &=&  \frac12 \Big[ \Phi^{q}_{i}(z,\zeta,s)
                  \pm \Phi^{q}_{i}(\ov{z},{\zeta},s)\Big] 
\qquad \mbox{~for~} i=A,P \,,
\end{eqnarray}
where we have used the symmetry relations (\ref{symm}).  With the same
relations one finds that $\Phi_V^{q(+)}$ and
$\smash{\Phi_{S,A,P}^{q(-)}}$ are odd under the replacement $\zeta
\leftrightarrow \ov{\zeta}$, which implies zeroes of these
distribution amplitudes at $\zeta=\frac12$.  Note also that the
$C$-even combinations $\Phi_{V,S}^{q(+)}$ are antisymmetric under the
replacement $z\leftrightarrow \ov{z}$ and therefore disappear in the
sum rules~(\ref{sumrules}).  This is consistent with the properties of
the form factors $F_V$ and $F_S$, which are $C$-odd.  The reverse
situation occurs for the $C$-odd combinations $\Phi_{A,P}^{q(-)}$,
which do not enter (\ref{sumrules}) in agreement with the $C$-even
nature of $F_A$ and $F_P$.

The distribution amplitudes $\Phi_i(z,\zeta,s)$ are complex
quantities, with phases due to the interactions in the $\bbb$ system.
Because of time reversal invariance they also parameterize matrix
elements with baryons in the initial state,
\begin{eqnarray} 
\lefteqn{P^+ \int \frac{dx^-}{2\pi}\, e^{i z P^+ x^-}\, 
\big\langle 0 \,\big|\, 
        \ov{q}(0)
            \gamma^+ q(\bar{x}) \,\big|\, 
   B(p\nu) \ov{B}(p'\nu') \,\big\rangle_{\rm in} = } 
\nn\\
&&\qquad\Phi^{q}_{V}(z,\zeta,s)\, \ov{v}(p'\nu') \gamma^+ \,u(p\nu)
       + \Phi^{q}_{S}(z,\zeta,s)\, \frac{P^+}{2m}\, \ov{v}(p'\nu') 
                          \,u(p\nu)\,, 
\nn\\
\lefteqn{P^+ \int \frac{dx^-}{2\pi}\, e^{i z P^+ x^-}\, 
\big\langle 0 \,\big|\, 
        \ov{q}(0) 
            \gamma^+\gamma_5\, q(\bar{x}) \,\big|\, 
   B(p\nu) \ov{B}(p'\nu')\, \big\rangle_{\rm in} =}  
\nn\\
&&\qquad\Phi^{q}_{A}(z,\zeta,s)\, \ov{v}(p'\nu') \gamma^+\gamma_5\, 
                                                            u(p\nu)
       - \Phi^{q}_{P}(z,\zeta,s)\, \frac{P^+}{2m}\,\ov{v}(p'\nu') 
                          \gamma_5\, u(p\nu) .
\end{eqnarray}
Notice the change of sign in front of $\Phi_P$.

We finally remark that distribution amplitudes for pairs $B_1
\ov{B}_2$ involving different baryons can be defined as in
(\ref{bbdaV}).  The charge conjugation relations (\ref{symm}) do not
hold for these quantities, but they do for distribution amplitudes of
the symmetrized states $| B_1(p\nu)\, \ov{B}_2(p'\nu') + B_2(p\nu)\,
\ov{B}_1(p'\nu') \rangle$.

\section{The handbag amplitude}
\label{sec:handbag}

We will now derive the expression for the soft handbag contribution to
$\gamma\gamma\to \bbb$.  The first steps of the derivation go in
complete analogy to the case of meson pair production.  We thus start
by summarizing the corresponding results of~\cite{DKV02}, and then
proceed from the point where the different nature of baryons and
mesons leads to important differences.  In~\cite{DKV02} we found an
appropriate frame to be the c.m.\ of the reaction, with axes chosen
such that the process takes place in the 1-3 plane and the outgoing
hadrons fly along the positive or negative 1-direction. Thus, we have
baryon momenta
\begin{equation}
p = \sqrt{\frac{s}{8}}
  \left[\, 1 \, , \, 1 \, , \, \sqrt{2}\beta\, {\bf e}_1 \,\right] , 
\qquad
p'= \sqrt{\frac{s}{8}}
  \left[\, 1 \, , \, 1 \, , \, -\sqrt{2}\beta\, {\bf e}_1 \,\right] , 
\end{equation}
with the relativistic velocity $\beta=\sqrt{1 - 4m^2 /s}$ and ${\bf
e}_1 = (1,0)$.  We hence have skewness $\zeta=\frac12$.  The photon
momenta read
\begin{eqnarray}
q &=& \sqrt{\frac{s}{8}}
      \left[\, 1+\sin\theta \, , \, 1-\sin\theta \, , 
            \, \sqrt{2} \cos\theta \, {\bf e}_1 \,\right] , 
\nonumber \\
q'&=& \sqrt{\frac{s}{8}}
      \left[\, 1-\sin\theta \, , \, 1+\sin\theta \, , 
            \, -\sqrt{2} \cos\theta \, {\bf e}_1 \,\right] ,
\end{eqnarray}
where $\theta$ is the c.m.\ scattering angle.  In terms of the usual
Mandelstam variables we have
\begin{equation}
 \cos\theta=\frac{t-u}{s}, 
\qquad 
 \sin\theta= \frac{2 \sqrt{t\, u}}{s}
\end{equation}
up to corrections of order $m^2 /s$.  The handbag amplitude for our
process can be written in terms of the hard scattering kernel for
$\gamma\gamma \to q\ov{q}$
\begin{eqnarray}
 H_{\alpha\beta}(k,k') = \left[ \epsilon \cdot \gamma \,
    \frac{(k-q)\cdot\gamma}{(k-q)^2+i \epsilon} \, \epsilon'\cdot\gamma
  +\epsilon'\cdot\gamma \, \frac{(q-k')\cdot\gamma}{(q-k')^2+i\epsilon} \,
     \epsilon\cdot\gamma \right]_{\alpha\beta}
\label{sub-amp}
\end{eqnarray}
with photon polarization vectors $\epsilon =\epsilon(q,\mu)$ and
$\epsilon'=\epsilon(q',\mu')$, and a matrix element describing the
$q\ov{q} \to \bbb$ transition.  Our starting expression thus is
\begin{eqnarray}
{\cal A}=\sum_q (e e_q)^2 \int\! d^4 k \,
    \int\! \frac{d^4 x}{(2 \pi)^4} \, e^{-i k\cdot x} \,
    {}_{\rm out}\big\langle B(p\nu) \, \ov{B}(p' \nu')\, \big| \,
    T \,\ov{q}{}_{\alpha}(x) \, q_{\beta}(0) \, \big|0 \big\rangle \;
  H_{\alpha\beta}(k,k') \,,
\label{handbag-amp}
\end{eqnarray}
where the summation index $q$ refers to the quark flavors $u$, $d$,
$s$.  In $H$ we have omitted terms suppressed by the current quark
masses.

As discussed in detail in \cite{DKV02}, the $q\ov{q}\to \bbb$
transition at large invariant mass $s$ can only be soft if the
incoming quark and antiquark have small virtualities and each carry
approximately the momentum of the baryon or antibaryon.  To quantify this,
we define $z = k^+ /P^+$ and parameterize the on-shell approximations
of the quark and antiquark momenta $k$ and $k'$ as
\begin{equation} 
\tilde{k}  = \sqrt{\frac{s}{2}}
   \left[\, z, \, \bar{z}, \, \sqrt{2z\bar{z}}\, \tr{e} \,\right] , 
\qquad 
\tilde{k}' = \sqrt{\frac{s}{2}}
   \left[\, \bar{z}, \, z, \, -\sqrt{2z\bar{z}}\, \tr{e} \,\right] ,
\end{equation}
where $\tr{e} = (\cos\varphi, \sin\varphi)$.  The requirements derived
in \cite{DKV02} then read
\begin{equation}
  2z-1, \, \sin\varphi \sim \frac{\Lambda^2}{s} \,,
\label{near-collinear}
\end{equation}
where $\Lambda$ is a hadronic scale of order 1~GeV.  In addition, the
minus- and transverse momenta of $k-\tilde{k}$ and $k' - \tilde{k}'$
must be of order $\Lambda^2/\sqrt{s}$.  As shown in \cite{DKV02,DFJK1}, 
the dominant Dirac structure of the soft matrix element in
(\ref{handbag-amp}) involves the good components of the quark fields
in the parlance of light-cone quantization.  Projecting these out we
have
\begin{equation}
{\cal A}_{\nu\nu',\mu\mu'} =
 - \sum_q (e e_q)^2  \int\! d^4 k \, \frac{1}{\sqrt{4 k^+ k'^+}}\,
\left[\, {\cal H}_{\mu\mu'}(\tilde{k},\tilde{k}')\, 
                                           {\cal S}_q(k,k')+ 
      {\cal H}^5_{\mu\mu'}(\tilde{k},\tilde{k}')\, 
                                           {\cal S}^5_q(k,k')\,
\right]  + {\cal O}(\Lambda^2/s) \,,
\label{handbag-3}
\end{equation}
where we have now made explicit the dependence on the photon and
baryon helicities $\mu$, $\mu'$ and $\nu$, $\nu'$, respectively.  Here
we have introduced the soft matrix elements\footnote{Note that we
define soft matrix elements for states with definite baryon momentum
here, and not for states $\langle {C+}|$ with definite charge parity
as in \protect\cite{DKV02}.}
\begin{equation}
{\cal S}_q = \frac{1}{2}
\int\! \frac{d^4 x}{(2 \pi)^4} \, e^{-i k\cdot x} \,
  {}_{\rm out}\big\langle \, B(p\nu) \ov{B}(p'\nu')
  \big| \, T \,\ov{q}(x) \, \gamma^+  q(0) \, \big|\, 0 \big\rangle \,,
\label{soft-mat}
\end{equation}
and ${\cal S}^5_q$ with $\gamma^+$ replaced by $\gamma^+\gamma_5$.
The hard subprocess amplitudes of $\gamma\gamma \to q\ov{q}$ for the
helicity sum and difference of the quark read
\begin{eqnarray}
{\cal H}_{\mu\mu'}(\tilde{k},\tilde{k}') 
   &=& \sum_{\lambda= \pm 1/2} \phantom{\lambda}
     \ov{u}(\tilde{k},\lambda) \, H_{\mu\mu'}(\tilde{k},\tilde{k}')\, 
             v(\tilde{k}',-\lambda)\,, 
\nn\\
{\cal H}^5_{\mu\mu'}(\tilde{k},\tilde{k}')
    &=& \sum_{\lambda= \pm 1/2} 2\lambda\; 
     \ov{u}(\tilde{k},\lambda) \, H_{\mu\mu'}(\tilde{k},\tilde{k}')\, 
             v(\tilde{k}',-\lambda)\,, 
\label{amp-com}
\end{eqnarray}
where we have approximated the parton momenta with their on-shell
values.  The expressions (\ref{handbag-3}) and (\ref{amp-com}) imply
the phase conventions for light-cone spinors given in the appendix.
With these conventions the behavior of helicity amplitudes under a
parity transformation is ${\cal A}_{\nu\nu',\mu\mu'}=\eta\,
(-1)^{\nu-\nu'-\mu+\mu'} {\cal A}_{-\nu-\nu',-\mu-\mu'}$, where $\eta$
is the product of the intrinsic parities of the four particles
involved.

According to our hypothesis that the $q\ov{q} \to \bbb$ transition is
dominated by soft processes, the matrix elements ${\cal S}_q(k,k')$
and ${\cal S}^5_q(k,k')$ should be strongly peaked
when~(\ref{near-collinear}) is fulfilled.  Depending on whether
$\varphi\approx 0$ or $\varphi\approx\pi$ this means that we have
$k\approx p$ or $k'\approx p$.  The case $k'\approx p$ corresponds to
the antiquark hadronizing into $B$ and the quark into $\ov{B}$.  This
requires sea quarks with very high momentum fraction in a baryon.  We
expect this to be disfavored compared with the case $k\approx p$, both
from phenomenological and theoretical considerations.  A rather direct
piece of information is for instance the ratio of quark and antiquark
distributions in a nucleon at large momentum fraction $x$.
Neglecting configurations with $k'\approx p$ compared with $k\approx
p$, we proceed by Taylor expanding ${\cal H}(\tilde{k},\tilde{k}')$
and ${\cal H}^5(\tilde{k},\tilde{k}')$ around $z=\frac{1}{2}$ and
$\varphi=0$, keeping only the leading order in $\Lambda^2/s$ .  We get
${\cal H}^{(5)}_{--} = {\cal H}^{(5)}_{++} = 0$ and
\begin{eqnarray}
{\cal H}_{+-}(\tilde{k},\tilde{k}')   
&=& \phantom{-} {\cal H}_{-+}(\tilde{k},\tilde{k}')
\:=\:  
   2\, \Big( \sqrt{ \hat{u}/\hat{t} } - \sqrt{ \hat{t}/\hat{u} } \,\Big)
\:=\:  2\, \frac{t - u}{\sqrt{tu}} + {\cal O}(\Lambda^2/s) ,
\nn\\[0.2em]
{\cal H}^5_{+-}(\tilde{k},\tilde{k}') 
&=& - {\cal H}^5_{-+}(\tilde{k},\tilde{k}')
\:=\:  
   2\, \Big( \sqrt{ \hat{u}/\hat{t} } + \sqrt{ \hat{t}/\hat{u} } \,\Big)
\:=\:  2\, \frac{s}{\sqrt{tu}} + {\cal O}(\Lambda^2/s) ,
\label{taylor}
\end{eqnarray}
where $\hat{t}$ and $\hat{u}$ are the Mandelstam variables of the hard
subprocess, with quark and antiquark momenta put on shell.  For better
legibility, explicit helicities are labeled only by their signs here
and in the following.  The amplitudes with equal photon helicities
$\mu=\mu'$ will be nonzero at next-to-leading order in $\als$, in
analogy to the photon helicity flip transitions in large-angle Compton
scattering \cite{hkm}.  With (\ref{taylor}) the loop integration in
(\ref{handbag-3}) now only concerns the soft matrix elements and leads
to moments of $\bbb$ distribution amplitudes.  With
$dk^+/\sqrt{4k^+k'{}^+}\simeq dz$ and the definitions of
Sect.~\ref{sec:gda} we obtain
\begin{eqnarray}
  \label{semi-final}
{\cal A}_{\nu\nu',\mu\mu'} &=&
 - \sum_q (e e_q)^2  \int_0^1\! dz  \, 
    \left\{{\cal H}_{\mu\mu'}(p,p')
     \,\Big[ \Phi_V^{q}\,(z,\zeta=\half,s)\,  \frac{1}{2P^+}\,
         \ov{u}(p\nu) \gamma^+ \,v(p'\nu')  \right.
\\[0.1em]
&&\hspace{12em} +\, 
          \Phi_S^{q}\,(z,\zeta=\half,s)\,
              \frac{1}{4m}\,\ov{u}(p\nu)\, v(p'\nu')\, \Big]
\nn\\[0.1em]
&&\hspace{7.1em} +\, {\cal H}^5_{\mu\mu'}(p,p')
     \,\Big[ \Phi_A^{q}\,(z,\zeta=\half,s)\, \frac{1}{2P^+}\,
         \ov{u}(p\nu) \gamma^+\gamma_5\, v(p'\nu') 
\nn\\[0.1em]
&&\hspace{12em} \left. +\,\Phi_P^{q}\,(z,\zeta=\half,s)\, 
           \frac{1}{4m}\,\ov{u}(p\nu)
                    \gamma_5\, v(p'\nu')\, \Big] \right\} 
\;+\; {\cal O}(\Lambda^2/s) \,.
\nn
\end{eqnarray}
Because of (\ref{symm}) the scalar distribution amplitude $\Phi_S$
decouples in our frame with $\zeta=\half$.  Evaluating the spinor
products with the conventions given in the appendix, including terms
suppressed only by $m/\sqrt{s}$, we arrive at our final result for the
$\gamma\gamma\to \bbb$ amplitudes:
\begin{eqnarray}
  \label{final}
\lefteqn{
{\cal A}_{\nu\nu',+-} 
= -(-1)^{\nu-\nu'}\, {\cal A}_{-\nu -\nu',-+} 
= 4\pi\ale\; \frac{s}{\sqrt{tu}}
}
\\
&\times& \left\{ 
        \delta_{\nu-\nu'}\, \frac{t-u}{s}\,  R_V(s)
    \;+\; 2\nu \delta_{\nu-\nu'} \Big[ R_A(s) + R_P(s) \Big]
    \;-\; \frac{\sqrt{s}}{2m}\, \delta_{\nu\nu'}\, R_P(s) \right\}
 \;+\; {\cal O}(\Lambda^2/s)\,,\nn 
\end{eqnarray}
where we have defined the annihilation form factors by
\begin{equation}
  R_{i}(s) = \sum_q e_q^2 \, F^q_{i}(s) \qquad
\mbox{for~} i=V,A,P \,, 
\label{moment}
\end{equation}
with $F^q_i$ from (\ref{sumrules}).  As in wide-angle Compton
scattering off baryons \cite{DFJK1} there are only three independent
form factors.  In Compton scattering it is the pseudoscalar rather
than the scalar form factor that does not contribute, due to different
choices of the reference frames.  The unpolarized differential cross
section is given by
\begin{equation}
 \frac{d\sigma}{d t}(\gamma\gamma \to \bbb) =
    \frac{4 \pi \ale^2}{s^2} \, \frac{1}{\sin^2\theta}\;
 \left\{\, \Big|R_{V}(s)\Big|^{\,2}\, \cos^2\theta \,
      + \,\Big|R_{A}(s) + R_P(s) \Big|^{\,2} \,
      + \,\frac{s}{4m^2}\, \Big|R_{P}(s)\Big|^{\,2} \,
 \right\}\, .
\label{dsdt-bbb}
\end{equation}

Several comments on our result (\ref{final}) are in order.  Unlike
$R_A$ and $R_P$, the vector form factor $R_V$ projects on the $C$ odd
part of the $\bbb$ state, whereas a $\gamma\gamma$ collision produces
of course its $C$ even projection.  This is a result of the
approximations that lead from (\ref{handbag-3}) to (\ref{semi-final}),
namely of neglecting configurations where the $\ov{q}$ emerging from
the two-photon annihilation hadronizes into the baryon $B$ instead of
the antibaryon $\ov{B}$.  To take this contribution into account, one
can split the loop integration over $\tr{k} = (k^1,k^2)$ into the two
hemispheres where $k^1$ is either positive or negative.  In the former
case one can expand the hard scattering amplitudes around $\varphi=0$, 
and in the latter around $\varphi=\pi$.  For $\varphi=\pi$ we then get
${\cal H}_{+-} \approx 2(u-t) /\sqrt{tu}$ instead of $2(t-u)
/\sqrt{tu}$ as in (\ref{taylor}).  The sum over both hemispheres thus
gives a result proportional to ${\cal H}_{+-}(p,p')$ times 
\begin{equation}
\int dk^- d^2\tr{k}\; \mbox{sgn}(k^1)\, {\cal S}_q(k,k') \,,
  \label{exact}
\end{equation}
instead of the integral
\begin{equation}
\int dk^- d^2\tr{k}\; {\cal S}_q(k,k') \,,
  \label{nice}
\end{equation}
which we used to express our result in terms of the distribution
amplitude $\Phi_V$.  The integrated matrix elements (\ref{exact}) and
(\ref{nice}) have opposite behavior under charge conjugation, but to
the extent that the region $k\approx p'$ gives a small contribution
compared to the region $k\approx p$, their difference can be
neglected.  In our derivation we have preferred the form (\ref{nice})
that leads to matrix elements of light-cone operators with well-known
properties, at the price of a loss in accuracy which we do not expect
to be critical.  We remark that for baryon helicities $\nu=\nu'$ both
(\ref{exact}) and (\ref{nice}) are odd under $z\leftrightarrow\ov{z}$
in our reference frame, which results in a zero contribution to the
amplitude after integration over $z$.  This can be shown by performing
a charge conjugation followed by a rotation of $180^\circ$ around the
$3$-axis.  The amplitudes with opposite baryon helicities $\nu= -\nu'$
do not decouple in this way, because charge conjugation exchanges the
helicities of $B$ and $\ov{B}$.  We finally emphasize that the
amplitude~(\ref{final}) for the $\gamma\gamma\to \bbb$ process does
have the correct behavior under charge conjugation, which for our
spinor convention reads ${\cal A}_{\nu\nu',\mu\mu'}=
(-1)^{\nu-\nu'-\mu+\mu'} {\cal A}_{\nu'\nu, \mu'\mu}$ in this channel.

For the sake of comparison let us mention what happens if we make the
corresponding approximations in wide-angle Compton scattering.  
Using that contributions where a fast antiquark is emitted from and 
reabsorbed by the baryon are small, one may count them with the ``wrong'' 
sign in the Compton form factors $R_V(t)$ and $R_T(t)$ of~\cite{DFJK1}. 
Replacing then the explicit factors $1/x$ with $1$, we obtain 
approximations $R_V(t) \approx \sum_q^{\phantom{2}} e_q^2\, 
F_1^q(t)$ and $R_T(t)\approx \sum_q^{\phantom{2}}  e_q^2\, F_2^q(t)$ 
in terms of the space-like Dirac and Pauli form factors, in analogy
with our result here. 

The amplitude (\ref{final}) shows important differences compared with
the one we obtained in \cite{DKV02} for production of a pair 
of pseudoscalar mesons.  Similarly to the $\bbb$ amplitudes with
$\nu=\nu'$, the matrix element corresponding to~(\ref{exact}) vanishes
there when integrated over~$z$.  Due to parity invariance the
corresponding contribution from ${\cal S}^5_q$ is also zero, so that
the leading term in the Taylor expansion (\ref{taylor}) of the hard
subprocess gives a vanishing scattering amplitude.  We thus had to
include the first nonleading term in this expansion, which is
proportional to $z-\ov{z}$.  If one were to do the same in the $\bbb$
case, one would get a further term in (\ref{final}).  It would involve
$\int dz (2z-1)\, \Phi_S(z,\frac{1}{2},s)$, which is a form factor of
the quark energy-momentum-tensor, in analogy to the meson pair case.
Such a subleading contribution would behave as $1/(tu)$ in the
amplitude instead of $1/\sqrt{tu}$, and give a $1/\sin^4\theta$ rather
than $1/\sin^2\theta$ dependence in the cross section.

Returning to our result (\ref{final}), we observe that the pseudoscalar
annihilation form factor $R_P$ generates amplitudes with equal
helicities of the baryon and antibaryon, i.e., with their spin
projections along ${\bf p}$ coupled to zero.  The spins of quark and
antiquark in the $q\ov{q}\to \bbb$ transition do then not sum up to
those of the $\bbb$ system, which implies that parton configurations
with non-zero orbital angular momentum along ${\bf p}$ are required in
$R_P$.  One expects that at large $s$ such quantities are suppressed
compared with $R_V$ or $R_A$.  An analog in the space-like region are
the form factors $R_T(t)$ vs.\ $R_V(t)$, and their electromagnetic
counterparts $F_2(t)$ vs.\ $F_1(t)$.  Recent measurements from
Jefferson Lab \cite{Gayou:2001qd} indicate that for $-t$ between 1~and
5.6~GeV$^2$ the ratio $F_2(t) / F_1(t)$ approximately scales as $m
/\sqrt{-t}$.  Assuming a similar behavior of $R_P(s) /R_A(s)$ one
finds that the term with $s |R_P(s)|^2$ in (\ref{dsdt-bbb}) will not
start do dominate over the other terms with increasing $s$.

We see in (\ref{dsdt-bbb}) that the form factor $R_V$ can be separated
from the two others through measurement of the angular distribution of
the $\bbb$ pair, given data of sufficient accuracy.  This requires
some lever arm in $\theta$, but must stay within the validity of our
approach, which is not applicable for $\theta$ near $0$ or $\pi$,
where the $\gamma\gamma\to q\ov{q}$ subprocess is no longer hard.  

A separation of $R_A$ and $R_P$ can only be performed with suitable
polarization measurements.  Our amplitudes (\ref{final}) are evaluated
for states with definite light-cone helicities, which is natural
within our framework and leads to simple expressions.  In the
unpolarized cross section this does not matter, but for polarization
observables the use of the ordinary c.m.s.\ helicity basis is more
convenient.  The transformation from one helicity basis to the other
can be found in~\cite{die2001}.  In our kinematics, the c.m.s.\
helicity amplitudes ${\cal M}$ read
\begin{equation} 
{\cal M}_{\nu\nu', \mu\mu'}\, = \, {\cal A}_{\nu\nu', \mu\mu'}\,
        +  \frac{m}{\sqrt{s}}\,
     \left[\,2\nu {\cal A}_{-\nu \nu',\mu\mu'}\,
          +\,2\nu' {\cal A}_{\nu -\nu',\mu\mu'}\,\right]
          \,+\, {\cal O}(m^2/s)\, .
\end{equation}
An observable capable of separating $R_P$ from the other form factors
is, for instance, the helicity correlation $C_{LL}$ between baryon and
antibaryon, given by
\begin{equation}
C_{LL}= \frac{d\sigma{(++)}-d\sigma{(+-)}}{d\sigma{(++)}+d\sigma{(+-)}}
      = -\, \frac{ 
   \Big|R_{A}(s)+ R_{P}(s) \Big|^{\,2} + 
                   \cos^2 \theta \Big|R_{V}(s)\Big|^{\,2} 
                 - \frac{s}{4m^2} \Big|R_{P}(s)\Big|^{\,2} }{
   \Big|R_{A}(s) + R_{P}(s) \Big|^{\,2}  + 
                   \cos^2 \theta \Big|R_{V}(s)\Big|^{\,2} 
                 + \frac{s}{4m^2} \Big|R_{P}(s)\Big|^{\,2} } \, ,
\end{equation}
where $d\sigma(\nu,\nu')$ is the cross section for polarized $\bbb$
production.

One may also consider the time-reversed process $\bbb\to
\gamma\gamma$, which for the case of proton-antiproton collisions may
be experimentally accessible and has already been mentioned in
\cite{ter01}.  Time reversal invariance relates the amplitudes of both
processes by
\begin{equation} 
{\cal A}_{\mu\mu',\,\nu\nu'}^{\bbb\to \gamma\gamma}\; (s,t) = 
  {\cal A}_{\nu\nu',\,\mu\mu'}^{\gamma\gamma\to \bbb}\; (s,t)\, .
\end{equation}
Up to an extra $(1-4 m^2/s)^{-1}$ in the phase space factor, $\bbb\to
\gamma\gamma$ has therefore the same cross section as $\gamma\gamma\to
\bbb$. The relation between the $\bbb$ distribution amplitudes for
baryons in the initial or in the final state has already been given in
Sect.~\ref{sec:gda}.  As in the case of wide-angle Compton scattering
\cite{DFJK2}, one can finally generalize our approach to the case of
virtual photons, provided their virtualities are at most of order $s$.

\section{Flavor symmetry}
\label{sec:flavor}

We are now going to discuss various $\bbb$ channels where $B$ is a
member of the lowest-lying octet of baryons.  We shall derive
relations among the various amplitudes and form factors in order to
simplify the analysis of experimental data on these cross section, and
to explore generic consequences of soft handbag dominance.

Relations among the various $\bbb$ channels are obtained by exploiting
flavor symmetry, i.e.\ isospin and $U$-spin invariance. The latter is
the symmetry under the exchange $d\leftrightarrow s$, and relates for
instance the $p\ov{p}$ and the $\Sigma^+\ov{\Sigma}{}^-$ channels.
Since the photon behaves as a $U$-spin singlet while $(p,\Sigma^+)$
and $(\ov{\Sigma}{}^-,\ov{p})$ are doublets, $U$-spin conservation
leads to
\begin{equation}
\rule[-0.5em]{0pt}{2em}
  {\cal A}(\Sigma^+ \ov{\Sigma}{}^-)\,\simeq\, {\cal A}(p\ov{p})\,.
\label{S-amp} 
\end{equation}
In contrast to isospin breaking, which is known to hold on the percent
level and will be neglected here, $U$-spin violations cannot
numerically be ignored.  This is indicated in (\ref{S-amp}) and in
later relations by the approximate symbol.  In analogy to
(\ref{S-amp}) one has
\begin{equation}
{\cal A}(\Xi^-\ov{\Xi}{}^+)\,\simeq\,  
                        {\cal A}(\Sigma^-\ov{\Sigma}{}^+)\,.  
\end{equation}
Other consequences of $U$-spin symmetry hold for the $U$-spin triplet
$\big(n, \half [\Sigma^0 + \sqrt{3}\Lambda], \Xi^0\big)$ and the
$U$-spin singlet $\half [\Lambda - \sqrt{3}\Sigma^0]$.  Together with
the corresponding transformation properties of the antiparticles one
obtains
\begin{eqnarray}
{\cal A}(\Xi^0\ov{\Xi}{}^0) &\simeq& {\cal A}(n\ov{n})
\hspace{0.6em}   \;\simeq\; \frac{1}{2}\, 
                   \Big[ {\cal A}(\Sigma^0\ov{\Sigma}{}^0)
                     - 3 {\cal A}(\Lambda\ov{\Lambda})\, \Big] \,,
\nonumber \\
{\cal A}(\Lambda\ov{\Sigma}{}^0) &\simeq& 
{\cal A}(\Sigma^0 \ov{\Lambda})
        \;\simeq\; \frac{\sqrt{3}}{2}\,
                   \Big[ {\cal A}(\Lambda\ov{\Lambda}) - 
                         {\cal A}(\Sigma^0\ov{\Sigma}{}^0)\, \Big] \,.
\end{eqnarray}
Notice that the preceding $U$-spin relations hold in any dynamical
approach respecting SU(3) flavor symmetry.

To proceed, we use that the handbag mechanism involves intermediate
$\qbq$ states,
\begin{equation}
\gamma\gamma \to \qbq \to \bbb \,,
\end{equation}
and generically decompose the $\gamma\gamma$ amplitudes as
\begin{equation}
{\cal A}(\bbb) = \sum_q e^2_q A_q^B\,.
\label{adecomp}
\end{equation}
We have omitted helicity labels since the following results hold for
any set of helicities.  The decomposition (\ref{adecomp}) already
follows from (\ref{handbag-amp}) and thus is more general than
our result (\ref{final}).  It does not rely on our neglect of various
${\cal O}(\Lambda^2/s)$ effects, nor of contributions where a fast
$\ov{q}$ hadronizes into a baryon.

A characteristic feature of the handbag mechanism is that the
intermediate $\qbq$ state can only be coupled to isospin $I=0$ and
$I=1$, but not to $I=2$.  This leads to particular strong restrictions
in the $\Sigma$ sector, where it reduces the number of independent
partial amplitudes $A_q^\Sigma$ to three, one for each flavor.  The
absence of the isospin-two component of $\Sigma \ov{\Sigma}$ implies
the following relation for the amplitudes:
\begin{equation}
{\cal A}(\Sigma^0\ov{\Sigma}{}^0) = -\frac12\, 
                   \Big[ {\cal A}(\Sigma^-\ov{\Sigma}{}^+)\,+\, 
                    {\cal A}(\Sigma^+\ov{\Sigma}{}^-)\, \Big]\,,
\end{equation}
which provides bounds on the (integrated or differential) 
$\Sigma^0\ov{\Sigma}{}^0$ cross section,
\begin{equation}
\frac12\,\left|\sqrt{\sigma(\Sigma^+\ov{\Sigma}{}^-)} -  
                \sqrt{\sigma(\Sigma^-\ov{\Sigma}{}^+}) \right| \leq
                \sqrt{\sigma(\Sigma^0\ov{\Sigma}{}^0)} \leq
\frac12\,\left(\sqrt{\sigma(\Sigma^+\ov{\Sigma}{}^-)} +
                \sqrt{\sigma(\Sigma^-\ov{\Sigma}{}^+)} \,\right)\,.
\end{equation}
This follows from isospin invariance alone and thus is a robust
prediction of handbag dominance.

Combining the relations due to isospin and to $U$-spin, in particular
the absence of final states with $I=2$ for the handbag, we find that
all $\bbb$ channels are described by only three independent partial
amplitudes, which one may take to be those of the $p\ov{p}$ channel,
$A^p_u$, $A^p_d$ and $A^p_s$.  The compiled relations of the other
partial amplitudes to the proton ones read
\begin{eqnarray}
A_u^n &=& A_d^p\,, \hspace*{2.7cm} A_d^n \;= A_u^p\,,
                   \hspace*{2.8cm} A_s^n\:\:\,=\: A_s^p\,, 
\phantom{\frac12}
\nn\\[0.3em]
A_u^{\Sigma^+}&=&A_d^{\Sigma^-}\simeq\, A_u^p\,, \hspace*{1.3cm} 
         A_d^{\Sigma^+}=A_u^{\Sigma^-}\simeq\, A_s^p\,, \hspace*{1.3cm} 
         A_s^{\Sigma^+}=A_s^{\Sigma^-}\simeq\, A_d^p\,,
\phantom{\frac12}
\nn\\[0.3em]
A_u^{\Sigma^0}&=& A_d^{\Sigma^0} \hspace{2pt}
                \simeq -\frac12\, [A_u^p +A_s^p]\,,  \hspace*{10.2em}
         A_s^{\Sigma^0}\,=\: -A_s^{\Sigma^+}\simeq\, -A_d^p\,,
\nn \\[0.3em]
A_u^\Lambda\;&=& A_d^\Lambda \hspace{7pt}
       \simeq\, {}- \frac16\, [A_u^p + 4A_d^p + A_s^p]\,, \hspace*{6.7em}
       A_s^\Lambda\:\, \simeq -\frac13\, [2A_u^p - A_d^p + 2A_s^p]\,,
\nn \\[0.3em]      
A_u^{\Lambda\ov{\Sigma}{}^0}&=& -A_d^{\Lambda\ov{\Sigma}{}^0} \;\simeq\;
     \hspace{3pt}
            \frac{1}{2\sqrt{3}}\, [A_u^p - 2A_d^p +A_s^p]\,,
     \hspace*{4.8em} 
     A_s^{\Lambda\ov{\Sigma}{}^0} =\, 0 \,,
\nn \\[0.3em]
A_u^{\Sigma^0 \ov{\Lambda}} &=& -A_d^{\Sigma^0 \ov{\Lambda}}\;\simeq\;
     \hspace{3pt}
     A_u^{\Lambda\ov{\Sigma}{}^0} \,, \hspace*{12.0em}
     A_s^{\Sigma^0 \ov{\Lambda}} =\, 0 \,,
\phantom{\frac12}
\nn \\[0.3em]
A_u^{\Xi^-}&=& A_d^{\Xi^0} \simeq\,  A_s^p\,, \hspace*{1.5cm}
         A_d^{\Xi^-}= A_u^{\Xi^0} \simeq\, A_d^p\,, \hspace*{1.3cm}
         A_s^{\Xi^-}= A_s^{\Xi^0} \simeq\, A_u^p\,.  
\phantom{\frac12}
\label{a:symm}
\end{eqnarray}

We now take recourse to valence quark dominance, which allows us to
use the amplitudes~(\ref{final}) and the form factors $F^{q,B}_i(s)$.
Valence quark dominance implies $F^{s,p}_i(s)=0$.  With this
simplification the symmetry relations (\ref{a:symm}) hold for the form
factors $F^{q,B}_i(s)$ as well, separately for $i =V, A, P$.  We
emphasize that in the context of the soft handbag amplitude, valence
quark dominance does \emph{not} assume that non-valence Fock states
are unimportant, since any number of soft partons with appropriate
quantum numbers can connect the two parton-hadron vertices in
Fig.~\ref{fig:handbag}b.  Rather, we neglect contributions from sea
quarks that carry almost all the momentum of a baryon, which should be
a good approximation.

With regard to the accuracy of the present data on
$\gamma\gamma\to\bbb$ we simplify further by taking a single value
$\rho$ for the $d/u$ ratio of all proton form factors $F_{V}^p$,
$F_{A}^p$, $F_{P}^p$,
\begin{equation}
F^{d,p}_i \,=\, \rho\, F^{u,p}_i \,, \qquad
i =V, A, P \,.
\label{udrel}
\end{equation}
For the numerical analysis we will perform in Sect.~\ref{sec:pheno}
this is not a severe restriction, since we find the annihilation cross
sections dominated by the sum of the axial and the pseudoscalar form
factor.  The parameter $\rho$ in~(\ref{udrel}) is then essentially the
one for the combination $F_A + F_P$.  For further simplicity we will
assume $\rho$ to be real-valued and independent of $s$ in our
analysis.  The ansatz $F_i^{d,p}(s) \propto F_i^{u,p}(s)$ parallels
the behavior of fragmentation functions for $d\to p$ and $u\to p$
transitions. The $d/u$ ratio in fragmentation is not well-known; a
value of $\rho = 1/2$ is for instance chosen in~\cite{kramer}.  For
time-like form factors one obtains the same value by analytic
continuation to the point $s=0$, where the Dirac form factors are
$F_{1}^{d,p}(0)=1$ and $F_{1}^{u,p}(0) =2$, if one makes the
assumption that their ratio does not change significantly between
$s=0$ and large time-like $s$. A value of $\rho= 1$ on the other hand
is suggested by the fact that both $u$ and $d$ are valence quarks of
the proton, and in order to produce the proton two quarks have to be
created from the vacuum in both cases. Still different, the Lund Monte
Carlo event generator~\cite{pythia} provides a value of only 0.25 for
leading protons (with $z >0.5$).

On the basis of this simple model for the soft physics input to the
handbag approach, we can write the $\bbb$ amplitudes as
\begin{equation}
{\cal A}(\bbb)\,=\, r_B\, {\cal A}(p\ov{p})\,,
\end{equation}
where $r_B=r_B(\rho)$.  The ratios of differential or integrated cross
sections are then determined by $r_B^2$. These factors read
\begin{eqnarray}
r_n\;\; &=& \frac{1+4\rho}{4+\rho} \,,
\nn\\[0.5ex]
r_{\Sigma^-} &\simeq& \frac{1+\rho}{4+\rho}\,,
\qquad\qquad\hspace{0.5em}
       r_{\Sigma^0} \simeq {}-\frac{1}{2}\,\frac{5+2\rho}{4+\rho}\,,
\qquad\qquad
       r_{\Sigma^+} \;\simeq\; 1 \,, \hspace{4em}
\nn\\[0.5ex]
r_{\Lambda}\, &\simeq& {}-\frac{3}{2}\, \frac{1+2\rho}{4+\rho}\,,
\nn\\[0.5ex]
r_{\Lambda\ov{\Sigma}{}^0} &\simeq& 
r_{\Sigma^0 \ov{\Lambda}} \;\simeq\; 
  \frac{\sqrt{3}}{2}\, \frac{1-2\rho}{4+\rho}\,,
\nn\\[0.5ex]
r_{\Xi^0} &\simeq& \frac{1+4\rho}{4+\rho}\,,
\qquad\qquad
       r_{\Xi^-} \simeq\; \frac{1+\rho}{4+\rho}\,.
\label{symm-rel}
\end{eqnarray}
We recall that these relations receive corrections from SU(3) flavor
symmetry breaking.  An investigation of the size and pattern of such
corrections is beyond the scope of this work.  We will see that
neglecting them at the present stage is justified by the
accuracy of available data.

\section{Comparison with experiment}
\label{sec:pheno}

A suitable and sufficiently precise set of data would allow for an
experimental determination of the annihilation form factors, quite
analogous to the measurement of electromagnetic form factors. As
already mentioned, the angular distribution of the $\bbb$ pair allows
one to separate $R^B_V$ from the combination of $R^B_A$ and $R^B_P$
given in (\ref{dsdt-bbb}).  The present data \cite{CLEOp}-\cite{L3} on
unpolarized cross sections (we exclude low-energy data from our study)
does not permit such detailed investigation.  Moreover, most data is
taken at energies which are rather low for the kinematic requirement
of large $s$, $t$, $u$ in the handbag approach. Below $\sqrt{s} \simeq
3 \gev$ the dynamics may be dominated by resonances.

It has long been known~\cite{bl1980} that for asymptotically large $s$
the process is amenable to a leading-twist QCD treatment, where the
transition amplitude factorizes into a hard scattering amplitude for
$\gamma\gamma \to q\bar{q}\, q\bar{q}\, q\bar{q} $ and a single-baryon
distribution amplitude for each baryon.  As already mentioned in the
introduction, the leading-twist result \cite{farrar} is way below the
experimental data.  This holds in particular if the single-proton
distribution amplitude is close to its asymptotic form under
evolution, for which there is growing evidence now \cite{bol96}.  In
view of this we consider that we make an acceptably small error in our
present work by altogether neglecting the leading-twist contribution
to the processes in question.  We remark that on the other hand the
diquark model, which is a variant of the leading-twist approach,
provides reasonable fits to the data, at least for the $p\ov{p}$
channel \cite{kro93}.  Notice that both the leading-twist and the
diquark approach give real-valued amplitudes in the $\gamma\gamma$
annihilation channel.  In contrast, the handbag approach makes no
generic prediction: the phase of the amplitude is determined by the
phases of the annihilation form factors, which may or may not be
small.

The annihilation form factors and the $\bbb$ distribution amplitudes
can presently not be calculated from first principles in QCD.
Contrary to generalized parton distributions, they do not admit a
direct representation as overlaps of light-cone wave functions
\cite{DFJK1,DFJK3} either.  Progress in describing generalized
distribution amplitudes within a Bethe-Salpeter approach has recently
been reported \cite{Tiburzi:2002mn}.  No model calculation is
currently available for the annihilation form factors in the $s$ range
where we need them.  We will therefore determine these form factors
phenomenologically.

Let us start with information from other processes.  The E760
collaboration \cite{E760} has measured the magnetic form factor of the
proton in the time-like region for $s$ in the range $8.9 \div 13.0\,
\gev^2$. For the scaled form factor a value of $s^2 |G_M^{\,p}| \simeq
3\, \gev\,^4$ has been found.  With (\ref{gm}), (\ref{moment}),
(\ref{udrel}) this implies for the annihilation form factor
\begin{equation}
s^2 \big| R_V^{\,p}(s) \big| 
       \approx  2.4\, \gev^4,\, 3\, \gev^4,\, 5 \, \gev^4
\qquad \mbox{for~} \rho=0.25,\, 0.5,\, 1 \,,
\label{est-RV}
\end{equation}
if we neglect the non-valence contribution from $F^{s,p}_V$.
As we already discussed, the form factor $R_P$ involves parton orbital
angular momentum. For lack of better information we estimate the
magnitude of $R_P$ by assuming
\begin{equation}
 \frac{\sqrt{s}}{2m_p}\,
\left|\, \frac{R_P^{\,p}(s)}{R_A^{\,p}(s)} \,\right| 
\;\approx\;
\frac{\sqrt{-t}}{2m_p}\; \frac{F_2^{p}(t)}{F_1^{p}(t)} 
\;\approx\; 0.37
\label{est-RP}
\end{equation}
for large $s$ and $t$, where the numerical value is from the
measurement \cite{Gayou:2001qd} of $F^p_2(t) /F^p_1(t)$ in the range
$-t = 1 \div 5.6 \gev^2$.

\begin{figure}[ht]
\begin{center}
\psfig{file=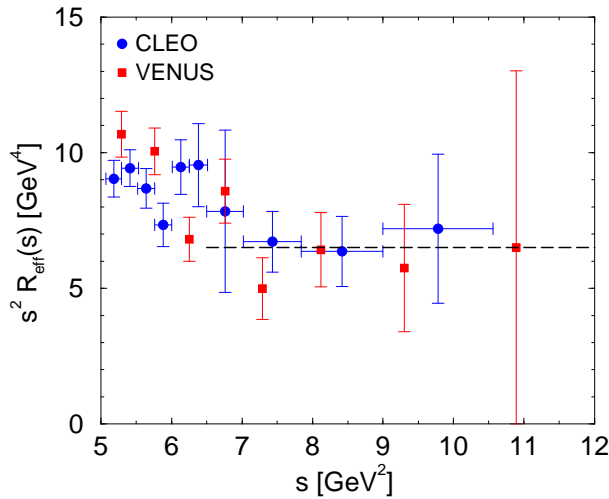,width=8.3cm}
\caption{The scaled annihilation form factor $s^2 R_{\rm eff}^{p}(s)$
as extracted from the data of \protect\cite{CLEOp,Hamasaki:1997cy},
using (\protect\ref{sig-special}) with $R_V^{\,p}$ set to zero.  The
dashed line represents a fit to the data above $6.5\,\gev^2$.}
\label{fig:rap}
\end{center}
\end{figure}

We now turn to the two-photon annihilation data.  Integrating the
cross section~(\ref{dsdt-bbb}) over $\cos\theta$ from $-\cos\theta_0$
to $\cos\theta_0$, we get
\begin{eqnarray}
\sigma(\gamma\gamma \to \bbb) &=& \frac{4 \pi \ale^2}{s}
 \left\{ \frac12 \ln\frac{1+\cos\theta_0}{1-\cos\theta_0} 
        \left(\,\Big|R_{V}^{B}(s)\Big|^{\,2}\, 
           +  \,\Big|R_{A}^{B}(s) + R_P^{B}(s) \Big|^{\,2} \,
           + \,\frac{s}{4m^2}\, \Big|R_{P}^{B}(s)\Big|^{\,2}\, \right)
\right.
\nn\\
  && \hspace{3.5em} \left.
     {}- \cos\theta_0\, \Big|R_{V}^{B}(s)\Big|^{\,2} \right\} \,.
\label{sig}
\end{eqnarray}
When comparing with data we need the integrated cross section for
$\cos\theta_0=0.6$, following the choice of the experiments,
\begin{equation}
 \sigma(\gamma\gamma \to \bbb) = 181 {\rm nb}\gev^2\, \frac{1}{s}\,
  \left\{ \big|R_A^{B} + R_P^{B}\big|^2 + \frac{s}{4m_B^2}\, \big|
  R_P^{B}\big|^2 + 0.134\, \big| R_V^{B}\big|^2 \right\} \,.
\label{sig-special}
\end{equation}
We fit this to the data on $\gamma\gamma\to p\bar{p}$ above
$s=6.5\,\gev^2 \approx (2.55\, \gev)^2$, trying to avoid as much as
possible the region where the process is markedly influenced by
resonances.  Such a fit determines the combination of form factors in
the curly brackets of (\ref{sig-special}).  Neglecting the term with
$R_V^{B}$ we get 
\begin{equation}
s^2 R^p_{\rm eff}(s) = (6.5 \pm 0.5) \gev^{\,4} \,,
\label{eq:reff}
\end{equation}
with the fit shown in Fig.~\ref{fig:rap}, where we have introduced the
abbreviation 
\begin{equation}
R_{\rm eff}^{B} = \left( \big|R_A^{B} + R_P^{B}\big|^2 
            + \frac{s}{4m_B^2}\, \big| R_P^{B}\big|^2 \right)^{1/2} .
\end{equation}
With our estimates (\ref{est-RV}) the contribution of $R_V^p$ to the
cross section (\ref{sig-special}) is at most 8\%.  Taking it into
account would thus reduce $R_{\rm eff}^p$ by at most $4\%$, which is
below the error in~(\ref{eq:reff}).  If we further use the estimate
(\ref{est-RP}) of $R^{\,p}_P /R^{\,p}_A$ at $s=6.5\,\gev^2$,
we obtain
\begin{equation}
  \label{eq:ra}
s^2 \big| R_A^{\,p}(s) \big| = (4.9 \pm 0.4) \gev^{\,4} \div
                               (8.0 \pm 0.6) \gev^{\,4} \,,
\end{equation}
where the errors are due to those in the fit (\ref{eq:reff}) and the
range to the uncertainty of the relative phase between $R_A$ and
$R_P$.  Using the same input we get the approximate relation $R_{\rm
eff}^p \approx |R_A^p + R_P^p|$, with an accuracy between 4\% and 11\%.

Although a behavior of $R_{\rm eff}^{p}(s) \sim s^{-2}$ is compatible
with experiment in the $s$ range we are investigating, a somewhat
different falloff is not excluded by present experimental data.  The
$s$ dependence of our fitted annihilation form factor coincides with
the one predicted by dimensional counting rules~\cite{bro:1973}, as
well as the corresponding behavior $d\sigma /dt \sim s^{-6}$ of the
cross section (\ref{dsdt-bbb}) at fixed angle $\theta$.  We emphasize
that this does not imply the dominance of leading-twist contributions.
It is also possible that, in a way similar to wide-angle Compton
scattering \cite{DFJK1,DFJK2}, dimensional counting rule behavior is
mimicked by soft physics over a large yet finite range of $s$.  {}From
our calculation of the handbag diagrams it is clear that the form
factors appearing in (\ref{final}) are only the \emph{soft} parts of
the matrix elements $R_{i}(s)$ ($i=V,\, A,\, P$) defined by
(\ref{timelike-ff}) and (\ref{moment}).  According to general power
counting arguments, the soft parts of $R_V$, $R_A$ and $\sqrt{s}\,
R_P$ will decrease faster than $s^{-2}$ for very large $s$.  The soft
handbag contribution to the cross section $d\sigma /dt$ then falls off
faster than $s^{-6}$ at fixed $\theta$, and the hard leading-twist
contribution will eventually dominate.  We remark that in the
spacelike region one can use a model based on wave function overlap to
evaluate the soft parts of the Compton form factors $R_V(t)$ and
$R_A(t)$ \cite{DFJK1}.  Their asymptotic behavior in this model is a
decrease like $t^{-4}$ and only sets in for $-t$ of order 100~GeV$^2$.

We observe that the annihilation form factor (\ref{eq:reff}) is of
similar size as the time-like magnetic form factor of the proton.  The
situation is thus similar to the space-like region, where the Dirac
form factor and the form factors for wide-angle Compton scattering off
the proton also behave similarly and are of comparable magnitude
\cite{DFJK1,DFJK2}.  Recall that the Compton form factors $R_A$ and
$R_V$ are given by moments in $x$ of generalized parton distributions
whose respective forward limits are the polarized and unpolarized
quark densities $\Delta q$ and $q$.  If one assumes that at large $x$
these generalized parton distributions have the same $x$-dependence as
their forward limits, up to a common factor $f(x,t)$ for both
distributions, one obtains $|R_A(t)| \le R_V(t)$ at large $t$ as a
consequence of the positivity bound $|\Delta q| \le q$.  For
generalized distribution amplitudes there is no such constraint, and
our estimates (\ref{est-RV}) and (\ref{eq:ra}) suggest that one may
indeed have $|R_A(s)| > |R_V(s)|$ for the annihilation form factors in
the $s$-range of our fit.

\begin{figure}[t]
\begin{center}
\psfig{file=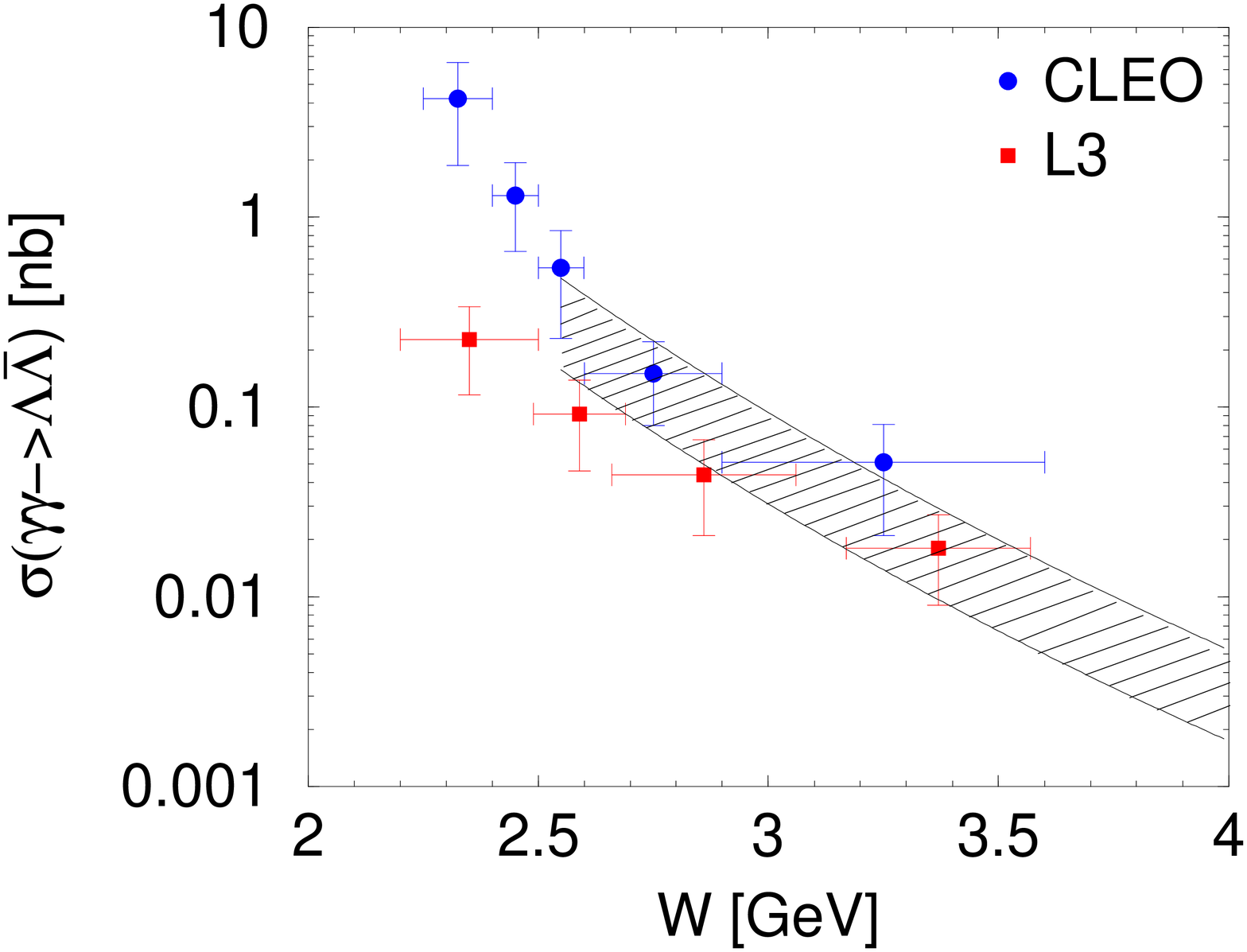,width=7.5cm}\hspace*{0.5cm}
\psfig{file=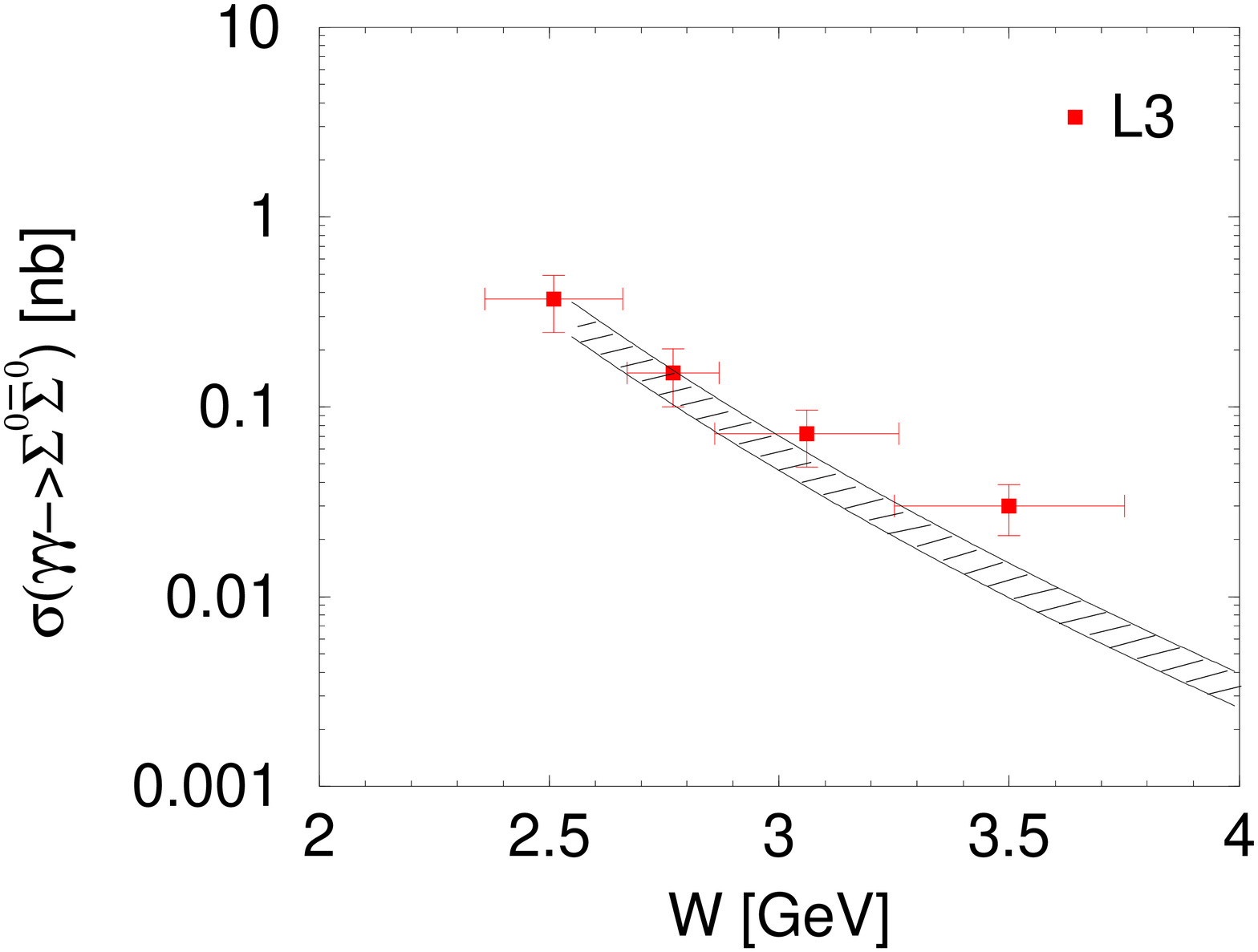,width=7.5cm}
\caption{The cross sections for two-photon annihilation into $\Lambda
\ov{\Lambda}$ (left) and $\Sigma^0 \ov{\Sigma}{}^0$ pairs (right)
versus $W=\sqrt{s}$. Data is taken from \cite{CLEOL,L3}. The bands
correspond to the range $\rho = 0.25 \div 0.75$ in conjunction with
the form factor $R_{\rm eff}^p(s)$ from (\protect\ref{eq:reff}), as
explained in the text.}
\label{fig:cross}
\end{center}
\end{figure}

Using the result (\ref{eq:reff}) for $R_{\rm eff}^{p}$ we can now
discuss the cross sections for other $B\ov{B}$ channels.  In view of
the large uncertainties of the data \cite{CLEOL,L3} we do not attempt
to include effects of flavor symmetry breaking and directly use the
relations (\ref{symm-rel}) to investigate the relative strength $\rho$
between $d\ov{d}\to p\ov{p}$ and $u\ov{u}\to p\ov{p}$ transitions.  In
Fig.~\ref{fig:cross} we show the cross sections for two-photon
annihilation into $\Lambda\ov{\Lambda}$ and $\Sigma^0\ov{\Sigma}{}^0$
pairs, with the bands corresponding to the range
\begin{equation}
\label{eq:rho}
  \rho = 0.25 \div 0.75 \,. 
\end{equation}
According to our discussion in Sect.~\ref{sec:flavor} such values are
physically quite plausible.  Values of $\rho$ significantly different
from (\ref{eq:rho}) are not favored by the data.  The estimate
(\ref{eq:rho}) should of course be interpreted with due care, given
theoretical uncertainties induced by the rather low $s$-values of the
data (the respective production thresholds for $\Lambda\ov{\Lambda}$
and $\Sigma^0\ov{\Sigma}{}^0$ pairs are at $\sqrt{s} = 2.23~\gev$ and
$2.39~\gev$), the assumed $s$-dependence of $R_{\rm eff}^{p}$ in
(\ref{eq:reff}), and the simplifying assumptions that give the
relations (\ref{symm-rel}) between $\bbb$ channels in terms of a
single real-valued parameter $\rho$.  We recall that the value in
(\ref{eq:rho}) essentially refers to the form factor combination
$R_{\rm eff}^{p} \approx |R_A^{\,p} + R_P^{\,p}|$, which dominates the
integrated cross section~(\ref{sig-special}).  Other $B\ov{B}$
channels can now easily be predicted from (\ref{symm-rel}).  A special
role is played by the mixed channels $\Lambda\ov{\Sigma}{}^0$ and
$\Sigma^0\ov{\Lambda}$, whose cross sections vanish at $\rho = 1/2$.
For these the estimate (\ref{eq:rho}) provides an upper bound
\begin{equation}
\sigma(\Lambda\ov{\Sigma}{}^0+ \Sigma^0\ov{\Lambda})
 \;\simeq\; \frac{3}{2}\, 
            \left(\frac{1-2\rho}{4+\rho}\right)^2  \sigma(p\ov{p})
 \;\le\; 0.02 \; \sigma(p\ov{p}) \,,
\end{equation}
whose precise value should again be taken with care, given our
discussion above.

\begin{figure}[ht]
\begin{center}
\psfig{file=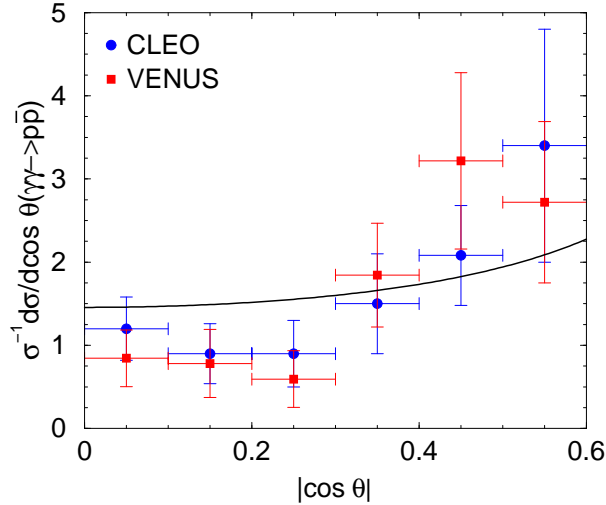,width=8.3cm}
\caption{Normalized angular distribution of $\gamma\gamma\to p\ov{p}$
at $s= 7.3 \gev\,^2$. The solid line represents the handbag result
(\protect\ref{dsdt-bbb}) with the form factor $R_V^{\, p}$ neglected.
Data is taken from \protect\cite{CLEOp,Hamasaki:1997cy}.}
\label{fig:ang-dis}
\end{center}
\end{figure}

In Fig.\ \ref{fig:ang-dis} we show the angular distribution for
$\gamma\gamma\to p\ov{p}$. Unfortunately, data exists only for rather
small energies, where our kinematical requirements that $-t$ and $-u$
should be large compared to, say, the squared proton mass, can hardly
be met.  Furthermore, the influence of resonances may not yet be
negligible, for which there might be a little hint in
Fig. \ref{fig:ang-dis}. At the energy of $\sqrt{s}\simeq 2.3 \gev$
(not shown in the figure) the data of~\cite{CLEOp,Hamasaki:1997cy}
exhibit a maximum at $90^\circ$, which is a clear signal for the
dominance of low partial waves and may be due to resonances.  The
comparison of the handbag result with the available data should
therefore be interpreted with due caution.  The curve in
Fig.~\ref{fig:ang-dis} shows the angular distribution of the handbag
result when $R_V^{\, p}$ is neglected in (\ref{dsdt-bbb}).  Taking our
estimate (\ref{est-RV}) of $R_V^{\, p}$ for $\rho = 0.5$, together
with the result~(\ref{eq:reff}) for $R_{\rm eff}^{p}$, we get a change
in the distribution that is too small to be seen in the figure.  If on
the other hand we take the value of $R_V^{\, p}$ which corresponds to
$\rho = 1$ in (\ref{est-RV}), the angular distribution becomes
somewhat steeper.  We emphasize however that the region where $-t$ or
$-u$ is smaller than $1.5\,\gev^2$ corresponds to $|\cos\theta| > 0.5$
for the data in Fig.~\ref{fig:ang-dis}.  In this region the handbag
result has to be taken with more than a grain of salt.

\section{Concluding remarks}

We have discussed the handbag contribution to two-photon annihilation
into baryon-antibaryon pairs at large energy and large momentum
transfer.  Our main result is to write the amplitude as a product of a
parton-level amplitude for $\gamma\gamma \to q\ov{q}$ and annihilation
form factors given by moments of the $B\ov{B}$ distribution
amplitudes.  In our derivation we have to explicitly neglect
contributions where the antiquark nearly takes the momentum of the
baryon and the quark the momentum of the antibaryon.  On the other
hand, quark off-shell effects in the hard scattering and the bad
components of the corresponding field operators are shown to be
suppressed parametrically.  An alternative treatment of the processes
under investigation is possible using double distributions
\cite{regen}.  Our results also apply to the annihilation process
$p\bar{p} \to \gamma\gamma$, whose form factors and amplitudes are
related to those for two-photon annihilation by time reversal.

The factorization of the soft handbag diagrams is analogous to the one
in wide-angle Compton scattering.  For the latter it has been shown
that this factorization remains valid when taking into account
next-to-leading corrections in $\als$ to the parton-level subprocess
\cite{hkm}, and one may expect that the same holds for the time-like
processes considered here.

The handbag contribution formally represents a power correction to the
leading-twist hard-scattering mechanism, but it seems to dominate at
experimentally accessible energies.  We find that the data for various
$\bbb$ channels is compatible with annihilation form factors
approximately behaving as $1/s^2$ for $s$ between $6$ and
$12\,\gev^{\,2}$, a counting rule behavior typical of many exclusive
observables.  Fitting the form factors to the data, we find that for
protons the sum of the axial and pseudoscalar annihilation form
factors $R_A + R_P$ is dominant and somewhat larger than the time-like
magnetic form factor.  A further test of our approach is the
approximate $1/\sin^2 \theta$ angular dependence of the cross section,
which agrees rather well with the VENUS data.  According to our
estimates, the $R_V$ term with its additional $\cos^2\theta$
dependences is likely too small to be seen in the presently available
data.  Flavor symmetry and the absence of $I=2$ components in the
$q\ov{q}$ intermediate states relate $p\ov{p}$ production to the
$B\ov{B}$ channels where $B$ is a member of the lowest lying baryon
octet, up to presumably moderate effects of flavor SU(3)
breaking. Fixing the relative strength $\rho$ of the form factors
governing $d\ov{d}\to p\ov{p}$ and $u\ov{u}\to p\ov{p}$ transitions
from suitable and sufficiently accurate data of two $B\ov{B}$ channels
allows one to predict all other ones.

We emphasize that our comparison with experiment suffers from the low
energies where data is currently available.  For these energies the
kinematical requirements of the handbag approach are hardly satisfied.
Nevertheless we arrive at a satisfactory description of the data for
the three channels $p\ov{p}$, $\Lambda\ov{\Lambda}$ and
$\Sigma^0\ov{\Sigma}{}^0$, taking as soft physics input the effective
form factor $R^p_{\rm eff}(s)$ and the flavor parameter $\rho$ with
values in agreement with the physical interpretation of these
quantities.  We finally remark that measurement of the process
$p\ov{p}\to \gamma\gamma$ with better statistics and at higher
energies would likely be possible at the proposed HESR project at
GSI~\cite{GSI}.

\section*{Note added}

After this work was finished, new data for the $p\bar{p}$ channel was
published by OPAL \cite{Abbiendi:2002bx}. In the high-$s$ range it agrees
with the results from CLEO and VENUS we have referred to in
Section~\ref{sec:pheno}.

\section*{Acknowledgments}  

We would like to thank Elliot Leader and Wolfgang Schweiger for
correspondence and Martin Siebel for providing us with information on
jet fragmentation into protons from the Lund Monte Carlo event
generator.  We also thank Michael D\"uren for his continued interest
in this topic.  This work is partially funded by the European
Commission IHP program under contract HPRN-CT-2000-00130.

\appendix
\section*{Appendix: Spinor conventions}
In our calculations we have used spinors for (anti)quarks and
(anti)baryons that correspond to states with definite light-cone
helicity~\cite{Kogut:1969xa}.  In the usual Dirac representation they
read
\begin{eqnarray}
  \label{spinors}
u(p,+) &=& N^{-1}
     \left( \begin{array}{c} 
      p^0 + p^3 + m \\ p^1 + i p^2 \\ p^0 + p^3 - m \\ p^1 + i p^2
     \end{array} \right) ,
\qquad \hspace{0.8em}
u(p,-) \:=\: N^{-1}
     \left( \begin{array}{c} 
      - p^1 + i p^2 \\ p^0 + p^3 + m \\ p^1 - i p^2 \\ - p^0 - p^3 + m
     \end{array} \right) ,
\nonumber \\
v(p,+) &=& N^{-1}
     \left( \begin{array}{c} 
      p^1 - i p^2 \\ - p^0 - p^3 + m \\ - p^1 + i p^2 \\ p^0 + p^3 + m
     \end{array} \right) ,
\qquad
v(p,-) \:=\: N^{-1}
     \left( \begin{array}{c} 
     - p^0 - p^3 + m \\ - p^1 - i p^2 \\ - p^0 - p^3 - m \\ - p^1 - i p^2
     \end{array} \right) ,
\end{eqnarray}
where $N = \sqrt{2(p^0 + p^3)}$.  This corresponds to the phase
conventions used by Brodsky and Lepage, cf.~\cite{Brodsky:1997de}, and
also to those of Kogut and Soper~\cite{Kogut:1969xa} if one takes into
account that they use a different representation of the Dirac
matrices.  The antiquark spinors in (\ref{spinors}) satisfy the charge
conjugation relations $v(p,\nu) = S(C)\, \ov{u}^T(p,\nu)$ with $S(C) =
i\gamma^2\gamma^0$.  For massless spinors one simply has $v(p,\nu) =
-u(p,-\nu)$.


\end{document}